\documentstyle[tighten,aps
]{revtex}

\newcommand{\be}{\begin{equation}}
\newcommand{\ee}{\end{equation}}
\newcommand{\lb}{\left (}
\newcommand{\rb}{\right )}

\twocolumn

\begin{document}

\preprint{TTP-99-24, hep-ph/9905553}

\date{May 1999}

\title{Virtual annihilation contribution to orthopositronium
decay rate }

\author{ Gregory Adkins\thanks{
  e-mail:  g$\_$adkins@acad.fandm.edu}}
  \address{Franklin and Marshall College, Lancaster, PA 17604}
\author{ Kirill Melnikov\thanks{
  e-mail:  melnikov@particle.physik.uni-karlsruhe.de}}
  \address{Institut f\"{u}r Theoretische Teilchenphysik,\\
            Universit\"{a}t Karlsruhe, D--76128 Karlsruhe, Germany}
\author{ Alexander Yelkhovsky\thanks{
  e-mail:  yelkhovsky@inp.nsk.su}}
  \address{ Budker Institute for Nuclear Physics,\\
            Novosibirsk, 630090, Russia}

\maketitle

\begin{abstract}
Order $\alpha^2$ contribution to the orthopositronium
decay rate due to one-photon virtual annihilation is
found to be  $\delta_{\rm ann}\Gamma^{(2)} =
        \lb \frac{ \alpha }{ \pi } \rb^2
        \lb \pi^2 \ln \alpha - 0.8622(9) \rb
        \Gamma_{\rm LO}$.

\vspace{0.3cm}


\end{abstract}
\vspace*{0.5cm}

Positronium, the bound state of an electron and positron,
is an excellent laboratory to test our understanding of
Quantum Electrodynamics of bound states.  Although in the majority of
cases the agreement between theory and experiment is very
good, the case of orthopositronium (o-Ps) decay 
into three photons
is outstanding, since the theoretical predictions differ
by 
about 6 standard deviations from the most accurate experimental
result \cite{expA} (see, however, an alternative result in \cite{expT}).
Provided that the experiment \cite{expA} is correct, the theory
can only be rescued if the second order correction to the o-Ps lifetime
turns out to be  $\sim 250 (\alpha/\pi)^2 \Gamma_{\rm LO}$.

It is however difficult to imagine how such a large number
could appear in the perturbative calculations, even
if the bound state is involved. This point of view is
supported by a recent  complete calculation of the ${\cal O}(\alpha^2)$
correction to the parapositronium (p-Ps) decay  rate into
two photons \cite{taupps}. It has shown that
the ``natural scale" of the gauge-invariant contributions is
[several units]$\times (\alpha/\pi)^2 \Gamma_{\rm LO}$.
For this reason, it was  conjectured in \cite{taupps} that
the ${\cal O}(\alpha^2)$ correction to the
orthopositronium decay rate
o-Ps $\to 3\gamma$ 
most likely is of the same order of magnitude.

At first sight, the result of Ref.\cite{AdLym},
\be
\delta_{\rm ann}\Gamma^{(2)} =
        \lb \frac{ \alpha }{ \pi } \rb^2
        \lb \pi^2 \ln \alpha + 9.0074(9) \rb
        \Gamma_{\rm LO},
\label{AL}
\ee
for the gauge-invariant contribution to the
o-Ps$\to 3\gamma$ decay rate induced by the single-photon virtual
annihilation, does not provide much support for this
conjecture.  In fact, the value of the non-logarithmic
constant in Eq.(\ref{AL}) is larger by approximately
one order of magnitude than the values of
coefficients in gauge-invariant contributions
to p-Ps$\to 2\gamma$ decay rate.

In this  note we would like to point out that
the result for the second order correction to
virtual annihilation contribution given in Eq.(\ref{AL})
is incomplete, in that closely related contributions should
be included as well. It turns out
that if the missing  pieces are added to Eq.(\ref{AL}), then
the complete result for $\delta_{\rm ann}\Gamma^{(2)}$
decreases and is in accord with the expectations advocated
in Ref. \cite{taupps}.

We recall, that in bound state calculations there are two
different types of contributions. The hard corrections
arise as contributions of virtual photons with momenta
$k \sim m$.  These contributions renormalize  local operators
in the non-relativistic Hamiltonian. For this reason they
can be computed without any reference to the bound state.

On the contrary, the soft scale contributions come from
a typical momenta scale $k \sim m\alpha$ in virtual loops,
and for this reason are sensitive to the bound state dynamics.
For $\delta_{\rm ann}\Gamma^{(2)}$, it is easy to see
that the soft scale contribution reads:
\be
\delta^{\rm (soft)}_{\rm ann}\Gamma^{(2)}  = \frac
{4\pi\alpha}{m^2} G(0,0),
\ee
where
$$
G(r,r') = \sum_n{\displaystyle{'}}\; \frac {|n(r) \rangle \langle n(r') |
}{E-E_n}
$$
is the reduced Green function of the Schr\"odinger equation
in the Coulomb field.

Let us write the expansion of the Green function in a series 
over the Coulomb potential:

\be
G(0,0) = G_0(0,0)+G_1(0,0) + G_{\rm multi}(0,0).
\ee
The first two terms in this expansion are divergent and require
regularization.  If we use dimensional regularization,
then the $G_0(0,0)$ piece delivers a finite contribution, since
it has only a power divergence. The second term $G_1(0,0)$
is logarithmically divergent. It can be easily seen that just
this term was accounted for in the calculation of
Ref.\cite{AdLym}, and it is precisely the term that
delivers the $\ln \alpha$ in Eq.(\ref{AL}). However, the
contributions of $G_0$ and $G_{\rm multi}$ were
not calculated there.

Both additional terms can be easily extracted from  Ref.\cite{levels}.
We then obtain
\be
\delta_{\rm 0}\Gamma =
        \frac{ 4\pi\alpha }{ m^2 } G_0(0,0) =
        \frac{1}{2} \alpha^2 \Gamma_{\rm LO}
\label{0C}
\ee
and
\be
\delta_{\rm multi}\Gamma =
        \frac{ 4\pi\alpha }{ m^2 }
        \sum_{n=2}^{\infty} G_n(0,0) =
        - \frac{3}{2} \alpha^2 \Gamma_{\rm LO},
\label{mC}
\ee
consistent with the results of Ref.\cite{Karsh}.

If we now add Eqs.(\ref{0C}), (\ref{mC}) and (\ref{AL}),
we obtain the complete ${\cal O}(\alpha^2)$ correction to
the o-Ps$\to 3\gamma$ decay rate due to single-photon
virtual annihilation:
\be
\delta_{\rm ann}\Gamma^{(2)} =
        \lb \frac{ \alpha }{ \pi } \rb^2
        \left \{  \pi^2 \ln \alpha - 0.8622(9) \right \}
        \Gamma_{\rm LO}.
\label{tot}
\ee

One sees that the non-logarithmic contribution is in fact of
order one times $(\alpha/\pi)^2 \Gamma_{\rm LO}$, in accord with
the conjecture in Ref.\cite{taupps}.
Its value is similar in magnitude to the known results for other
gauge-invariant ${\cal O}(\alpha^2)$ corrections to the
decay rate of orthopositronium \cite{Karsh,tauops}.

The only known exception from this ``rule" is provided by
the square of the ${\cal O}(\alpha)$ corrections
to the o-Ps decay amplitude. This (gauge-invariant)
contribution has an anomalously large coefficient:
$28.860(2)(\alpha/\pi)^2 \Gamma_{\rm LO}$ \cite{A2}. In this respect,
we would like to note that there may be an enhancement
factor due to a larger (by about a factor of 3) number of
diagrams contributing to o-Ps decay as compared to p-Ps decay.
This enhancement is seen already in the magnitude of the
${\cal O}(\alpha)$ corrections and translates naturally to the
large value of the ${\cal O}(\alpha ^2)$ contribution
originating from the square of the ${\cal O}(\alpha)$ corrections
to the o-Ps decay amplitude. However, unless this enhancement
is dramatic, it is hard to believe that this fact alone
can explain the discrepancy between theoretical and experimental
results on o-Ps decay rate.

This research was supported in part by the National Science
Foundation under grant number PHY-9722074, by BMBF under
grant number BMBF-057KA92P, by Gra\-duier\-ten\-kolleg
``Teil\-chen\-phy\-sik'' at the University of Karlsruhe,
by the Russian Foundation for Basic Research under grant number
99-02-17135 and by the Russian Ministry of Higher Education.

\end{document}